\documentclass[twocolumn,fleqn,natbib]{svjour2}
\bibpunct{[}{]}{;}{n}{}{,} 
\smartqed  
\usepackage{graphicx}
\usepackage{bm}
\journalname{Granular Matter}
\begin{document}

\title{Average energy and fluctuations of a granular gas in the threshold of the clustering instability}


\author{J. Javier Brey   \and M.J. Ruiz-Montero}


\institute{J.J. Brey \at
              Fisica Te\'{o}rica, Universidad de Sevilla, Apartado de Correos 1065. 41080 Sevilla. Spain \\
              Tel.: +34-954550936\\
              \email{brey@us.es}            \\  \and
           M.J. Ruiz-Montero \at
           Fisica Te\'{o}rica, Universidad de Sevilla, Apartado de Correos 1065. 41080 Sevilla. Spain \\
           \email{majose@us.es}
           }

\date{Received: date }

\maketitle

\begin{abstract}
The behavior of an isolated dilute granular gas near the threshold
of its clustering instability is investigated by means of
fluctuating hydrodynamics and the direct simulation Monte Carlo
method. The theoretical predictions from the former are shown to be
in good agreement with the simulation results. The total energy of
the the system is renormalized by the fluctuations of the vorticity
field. Moreover, the scaled second moment of the energy fluctuations
exhibits a power-law divergent behavior

\end{abstract}

\section{Introduction}
\label{s1} Granular materials often exhibit flows that are similar
to those found in molecular fluids. In fact, phenomenological
hydrodynamic-like equations are frequently used to describe them
\cite{Ha83}. Nevertheless, in general, fluctuations of the
hydrodynamic fields are much larger in granular systems than in
ordinary fluids \cite{GSyS04}, as a consequence of the number of
particles being many orders of magnitude smaller in the former than
in the latter. Thus fluctuations are expected to play a more
significant role in granular flows than in molecular fluid flows. In
addition, it must be realized that a new source of noise and
fluctuations is present in granular systems, associated with the
localized character of the energy dissipation. The properties of
this intrinsic noise remain largely unknown, although some simple
limiting cases have been recently investigated
\cite{BGMyR04,VPBWyT06}.

Here, some results for the total energy fluctuations in a freely
evolving granular gas, below the critical size for the onset of the
clustering or shear instability \cite{GyZ93}, will be reported. The
model considered is a system of $N$ smooth inelastic hard spheres of
mass $m$ and diameter $\sigma$. The inelasticity is assumed to be
described by a constant, velocity independent, coefficient of normal
restitution $\alpha$. It is well known that the simplest possible
state for this system is the so-called homogeneous cooling state
(HCS). At the level of hydrodynamics, this state is described by a
constant number density $n_{h}$, a vanishing flow velocity, and a
time dependent temperature $T_{h}(t)$, that obeys the equation
\begin{equation}
\label{1.1}
\partial_{t} T_{h}(t) = - \zeta_{h}(t)T_{h}(t),
\end{equation}
where $\zeta_{h}[n_{h},T_{h}(t)]$ is a cooling rate that must be
specified from a microscopic theory. For the case of hard spheres
considered here, there is no microscopic energy scale associated
with the collision model. Therefore, the temperature dependence of
the cooling rate can be determined by dimensional arguments as
$\zeta_{h}(t) \propto T_{h}^{1/2}(t)$. The HCS is unstable against
long wavelength spatial perturbations, leading to the formation of
velocity vortices and high density clusters of particles
\cite{GyZ93}. Linear stability analysis of the hydrodynamic
equations indicates that this instability is driven by the
transversal shear mode \cite{GyZ93,BRyC99}. Using phenomenological
Navier-Stokes equations, it is found that the critical wavelength
$L_{c}$ beyond which the system becomes unstable is given by
\begin{equation}
\label{1.2} L_{c}= 2 \pi \left( \frac{2 \eta^{*}}{\zeta^{*}}
\right)^{1/2} \ell_{0}.
\end{equation}
Here, $\ell_{0} \equiv (n_{h} \sigma^{2})^{-1}$ is proportional to
the mean free path,
\begin{equation}
\label{1.3} \zeta^{*} \equiv \frac{\zeta_{h}(t)
\ell_{0}}{v_{h}(t)},
\end{equation}
and
\begin{equation}
\label{1.4} \eta^{\ast} \equiv \frac{\eta(T_{h})}{m n_{h} \ell_{0}
v_{h}(t)}\, ,
\end{equation}
where $v_{h}(t) \equiv \left[2T_{h}(t)/ m \right]^{1/2}$ is a
thermal velocity and $\eta(T_{h})$ is the shear viscosity of the
granular fluid. For the particular case of a dilute gas described
by the inelastic Boltzmann equation, which is the case to be
considered in the following, the explicit expressions of $\eta$
and $\zeta_{h}(t)$ are given in \cite{BDKyS98}.

Although the initial set up of the clustering instability has been
extensively studied, much less attention has been payed to the
behavior of the system as the instability is approached from below,
i.e. as the linear system size $L$ increases towards $L_{c}$. The
aim of this paper is to analyze the behavior of the total energy of
of a dilute granular gas of hard spheres in that region, by using
fluctuating hydrodynamics and the direct simulation Monte Carlo
(DSMC) method \cite{Bi94}. A similar study for a system of hard
disks, where the simulations were carried out by means of molecular
dynamics, has already been reported elsewhere
\cite{BGMyR05,BDGyM06}.

\section{Fluctuating hydrodynamics and average energy}
\label{s2}

One of the main advantages of the DSMC method \cite{Bi94} is that it
allows to explote the symmetry of the state to be investigated, when
dividing the system into cells to apply the algorithm. For instance,
to study average properties of the HCS, the system can be forced to
stay homogeneous by considering a single cell, the position of the
particles then becoming irrelevant. This allows to significantly
increase the statistics of the results, decreasing the uncertainty.
Of course, by doing this all the spatial hydrodynamic fluctuations
and correlations that may occur in the system are by definition
eliminated. This implies, in particular, that the clustering
instability can not develop and its influence on the properties of
the system can not be studied with a single cell.

Nevertheless, it is possible to consider intermediate situations
where the clustering instability is present, but the system is
forced to have some kind of symmetry that allows to increase the
statistical accuracy. This can be achieved, in particular, by
dividing the system into a given number of parallel layers of the
same width. Each layer is considered as a cell when applying the
DSMC algorithm. Since the positions of particles inside the same
cell play no role in determining their collision probability, it
follows that the dynamics is somehow coarse-grained over each layer
and variations of the properties inside it can not be analyzed. On
the other hand, variations of the hydrodynamic fields between
different layers, i.e. along the coordinate perpendicular to them,
can show up. Therefore, the clustering instability appears when the
length of the system along that direction, $L_{x}$, reaches the
critical value $L_{c}$. This is the coarse-grained description for
which the theory will be developed in the following.

The total energy $\widetilde{E}(t)$ of the system, either isolated
or with periodic boundary conditions, we are considering can be
expressed as
\begin{eqnarray}
\label{2.1} \widetilde{E}(t)  & \equiv & \sum_{i=1}^{N} \frac{m
V_{i}^{2}(t)}{2} \nonumber \\
& = & \int dx \left[ \frac{3}{2}
\widetilde{N}_{x}(x,t)\widetilde{T}(x,t)+\frac{m}{2}
\widetilde{N}_{x}(x,t) \widetilde{u}^{2}(x,t) \right], \nonumber \\
\end{eqnarray}
where ${\bm V}_{i}(t)$ is the velocity of particle $i$ at time $t$,
the tilde is used to indicate that the quantities are treated as
stochastic variables, and the $x$ axis has been taken perpendicular
to the layers described above. In the equation,
$\widetilde{N}_{x}(x,t)$ is the number of particles per unit of
length in the $x$ direction, $\widetilde{{\bm u}}(x,t)$ is the flow
velocity field, and $\widetilde{T}(x,t)$ the temperature. The last
two quantities are coarse-grained over the cells. The microscopic
definitions of these fields are
\begin{equation}
\label{2.2} \widetilde{N}_{x}(x,t)=  \sum_{i=1}^{N} \delta \left[ x-
X_{i}(t) \right],
\end{equation}
\begin{equation}
\label{2.3} \widetilde{N}_{x}(x,t) \widetilde{\bm u}(x,t) =
\sum_{i=1}^{N} {\bm V}_{i}(t) \delta \left[ x- X_{i}(t) \right],
\end{equation}
\begin{eqnarray}
\label{2.4} \frac{3}{2} \widetilde{N}_{x}(x,t) \widetilde{T}(x,t) &
= &  \frac{m}{2} \sum_{i=1}^{N} \left[ {\bm V}_{i}(t)
-\widetilde{{\bm u}}(x,t) \right]^{2} \delta \left[ x- X_{i}(t)
\right], \nonumber \\
\end{eqnarray}
that correspond to the idealized limit of infinitely narrow layers.
Note that, consistently with the discussion above, only the $x$
coordinate of the particles at time $t$, $X_{i}(t)$, appears in the
definition of the coarse-grained fields. Expansion of Eq.\
(\ref{2.1}) around the average values of the fields in the HCS,
$Y_{\alpha,h}$, retaining up to second order terms in the
deviations, $\delta \widetilde{Y}_{\alpha}(x,t) \equiv Y_{\alpha,h}-
\widetilde{Y}_{\alpha}(x,t)$, yields
\begin{eqnarray}
\label{2.5} \delta \widetilde{E}(t) & \equiv & \widetilde{E}(t) -
E_{h}(t) \nonumber \\
& = & \frac{1}{2} \int dx\, \left\{ 3 N_{x,h} \delta
\widetilde{T}(x,t)+3 \delta \widetilde{N}_{x}(x,t) \delta
\widetilde{T}(x,t) \right. \nonumber \\
& & + \left. m N_{x,h} \left[ \delta \widetilde{\bm u}(x,t)
\right]^{2} \right\},
\end{eqnarray}
with $E_{h}(t) \equiv 3 N T_{h}(t)/2$ and $N_{x,h} \equiv N/L_{x}$.

In order to eliminate from the analysis the time dependence of the
reference HCS, it is convenient to introduce dimensionless length
$\ell$ and time $s$ scales by
\begin{equation}
\label{2.6} \ell = \frac{x}{\ell_{0}}, \quad ds
 = \frac{v_{h}(t)dt}{\ell_{0}}\, .
\end{equation}
Also, dimensionless hydrodynamic fields are defined as
\begin{equation}
\label{2.7} \rho_{x}(\ell,s) \equiv \frac{\delta
\widetilde{N}_{x}(x,t)}{N_{x,h}},
\end{equation}
\begin{equation}
\label{2.8} {\bm \omega}(\ell,s) \equiv \frac{\delta \widetilde{\bm
u}(x,t)}{v_{h}(t)},
\end{equation}
\begin{equation}
\label{2.9} \theta (\ell,s) \equiv \frac{\delta
\widetilde{T}(x,t)}{T_{h}(t)}.
\end{equation}
Moreover, the Fourier transform is introduced through
\begin{equation}
\label{2.10} \rho_{x,k}(s) = \int d \ell\,  e^{-i k \ell}
\rho_{x}(\ell,s),
\end{equation}
and similarly for the other fields. Then, Eq.\ (\ref{2.1}) becomes
\begin{eqnarray}
\label{2.11} \epsilon(t) & \equiv & \frac{\delta
\widetilde{E}(t)}{E_{h}(t)} \nonumber \\
& = & \frac{\theta_{0}(s)}{\Lambda_{x}} + \frac{1}{\Lambda_{x}^{2}}
\sum_{k} \left[ \rho_{x,k}(s) \theta_{-k}(s) + \frac{2}{3} |{\bm
\omega}_{k}(s)|^{2} \right],
\end{eqnarray}
with $\Lambda_{x} \equiv L_{x}/\ell_{0}$. Then, to calculate
$\epsilon(t)$, expressions for the Fourier components of the
fluctuating hydrodynamic fields are needed. It will be assumed that,
at the mesoscopic level used here, they obey Langevin equations
obtained by linearizing the Navier-Stokes equations for a granular
gas around the HCS. Moreover, if attention is restricted to the
quasi-elastic limit, i.e. $\alpha$ very close to unity, it can be
expected that a good approximation be obtained by using the same
expressions for the properties of the noise terms in the Langevin
equations as those in the Landau-Langevin equations for normal
fluids \cite{LyL59,vNEByO97}.

Consider the Fourier transform of the flow field, ${\bm
\omega}_{k}(s)$. The vorticity field or transverse flow field in a
general situation is by definition its component perpendicular to
the vector ${\bm k}$, i.e. in the present case perpendicular to
the $x$ axis. It will be denoted by ${\bm \omega}_{k,\perp}(s)$.
The coarse-grained velocity field ${\bm \omega}(\ell,s)$ is
related with the actual velocity field ${\bm w}({\bm l},s)$ by
\begin{equation}
\label{2.12} \int d{\bm \ell}_{\perp}\, {\bm w}({\bm \ell},s)=
\frac{V^{*}}{\Lambda_{x}} {\bm \omega}(\ell,s),
\end{equation}
where ${\bm \ell}={\bm r}/\ell_{0}$, ${\bm \ell}_{\perp}$ denotes
the vector component of ${\bm \ell}$ perpendicular to the $x$
axis, and $V^{*}$ is the volume of the system in the dimensionless
scale, $V^{\ast}=V/ \ell_{0}^{3}$. Using this, it is easily
obtained that the Landau-Langevin equation for ${\bm
\omega}_{k,\perp}(s)$ is
\begin{equation}
\label{2.13} \left( \frac{\partial}{\partial s}
-\frac{\zeta^{\ast}}{2}+\eta^{*} k^{2} \right) {\bm
\omega}_{k,\perp} (s)= {\bm \xi}_{k,\perp}^{(\omega)}(s).
\end{equation}
The term ${\bm \xi}_{k,\perp}^{(\omega)}(s)$ is a Gaussian white
noise verifying the fluctuation dissipation relation
\begin{equation}
\label{2.14} < {\bm \xi}^{(\omega)} _{k,\perp} {\bm
\xi}^{(\omega)}_{k^{\prime},\perp}(s^{\prime})>=\frac{\Lambda_{x}^{2}}{N}
\delta (s-s^{\prime}) \delta_{k,-k^{\prime}} \eta^{\ast} k^{2} {\sf
I},
\end{equation}
${\sf I}$ being the unit tensor of dimension $2$ and the angular
brackets denoting average over the noise realizations. It is worth
to remark that the clustering instability manifests itself in Eq.\
(\ref{2.13}), which shows that for $|k|<k_{c}\equiv \left(
\zeta^{\ast}/2 \eta^{\ast} \right)^{1/2}$, the fluctuations of the
scaled transverse flow field grow in time. On the other hand, for
$\Lambda_{x} < \Lambda_{c}=2 \pi /k_{c}$, the long time solution of
Eq.\ (\ref{2.13}) is
\begin{equation}
\label{2.15} {\bm \omega}_{k,\perp}(s)= \int_{-\infty}^{s}
ds^{\prime} e^{(s-s^{\prime})\lambda_{\perp}(k)} {\bm
\xi}_{k,\perp}^{(\omega)}(s^{\prime}),
\end{equation}
with $\lambda_{\perp}(k)=\frac{\zeta^{*}}{2}-\eta^{*} k^{2}$. Then,
by using Eq.\ (\ref{2.14}) it is obtained:
\begin{eqnarray}
\label{2.16} <{\bm \omega}_{k,\perp}(s) {\bm
\omega}_{k^{\prime},\perp}(s^{\prime})> & = &
-\frac{\Lambda_{x}^{2}\eta^{\ast} k^{2}}{2N \lambda_{\perp} (k)}
e^{(s-s^{\prime})\lambda_{\perp}(k)} \delta_{k,-k^{\prime}} {\sf
I} \nonumber \\
\end{eqnarray}
and, in particular,
\begin{equation}
\label{2.17} <|{\bm \omega}_{k \perp} (s) |^{2}> =
-\frac{\Lambda_{x}^{2}\eta^{\ast} k^{2}}{N \lambda_{\perp} (k)}.
\end{equation}
Equation (\ref{2.16}) holds for $s > s^{\prime} \gg 1$.

A Langevin equation for the energy is obtained by linearizing the
macroscopic average equation for it,
\begin{equation}
\label{2.18} \frac{d}{dt}\, \delta \widetilde{E}(t) = -\frac{9}{4}
\zeta_{h}(T_{h}) N_{x,h} \int dx\, \delta \widetilde{T} (x,t).
\end{equation}
For a normal fluid, the right hand side identically vanishes since
the cooling rate is zero. Then, an equation for the scaled energy
fluctuations is easily found,
\begin{equation}
\label{2.19} \frac{d \epsilon (s)}{ds} -\zeta^{\ast} \epsilon (s)
=-\frac{3 \zeta^{\ast}}{2 \Lambda_{x}}\ \theta_{0}(s).
\end{equation}
The noise term mentioned above, intrinsic to the local energy
dissipation in collisions, has been omitted, since it is expected to
give small contributions as compared with those to be kept, in the
quasi-elastic limit. Combination of Eqs.\, (\ref{2.11}) and
(\ref{2.19}) yields
\begin{eqnarray}
\label{2.20} \frac{ d \epsilon (s)}{d s} &=& - \zeta^{\ast}
\left\{ \frac{\epsilon(s)}{2} -\frac{3}{2 \Lambda_{x}^{2}} \right.
\nonumber \\
& & \left. \times \sum_{k} \left[ \rho_{k,x}(s) \theta_{-k}(s)
+\frac{2}{3} |{\bm \omega}_{k, \perp}|^{2} \right] \right\},
\end{eqnarray}
In the threshold of the clustering instability, the fluctuations
of the transversal components of the flow velocity are expected to
dominate over the density and the longitudinal velocity
fluctuations, so the above equation can be reduced to
\begin{equation}
\label{2.21} \frac{ d \epsilon (s)}{d s} = -
\frac{\zeta^{\ast}}{2}  \left[ \epsilon(s) -\frac{2}{
\Lambda_{x}^{2}} \sum_{k}  |{\bm \omega}_{k, \perp}|^{2} \right].
\end{equation}
>From this equation, it follows that the average value of the total
energy of the system is
\begin{equation}
\label{2.22} < \epsilon >_{st}  =  \frac{2}{\Lambda_{x}^{2}}
\sum_{k} \langle | {\bm \omega}_{k, \perp}(s)|^{2} \rangle_{st}   =
-\frac{2}{N} \sum_{k} \frac{\eta^{\ast}k^{2}}{\lambda_{\perp}(k)}\,
.
\end{equation}
This is the result expected from the expression obtained for a
system without introducing the coarse-grain average over the
parallel layers \cite{BDGyM06}.

For $\widetilde{\delta L} \equiv (L_{c}-L_{x})/L_{c} \ll 1$ and
positive, the main contribution to Eq.\ (\ref{2.22}) is given by
those modes having the largest possible wavelength, i.e. those with
$|k|=k_{min}=2 \pi/\Lambda_{x}$. For them, it is
\begin{eqnarray}
\label{2.23} \lambda_{\perp}(k) & = & \lambda_{\perp}(k_{min}) =
\frac{\zeta^{*}}{2} -\eta^{*} \left( \frac{ 2\pi}{\Lambda_{x}}
\right)^{2} \nonumber \\
& = & \frac{\zeta^{*}}{2} \left[ 1- \left( \frac{L_{c}}{L_{x}}
\right)^{2} \right] \simeq -\zeta^{\ast} \widetilde{\delta L}.
\end{eqnarray}

It is at this point when the coarse-grain used in the simulations
deserves some attention. In a normal simulation, for instance using
molecular dynamics of a cubic cell, the number of modes with
$k=k_{min}$ compatible with the periodic boundary conditions should
be $6$, two for each of the axis of the system \cite{BDGyM06}.
Nevertheless, the coarse-grain in the directions perpendicular to
the layers, actually kills all the modes in those directions.
Consequently only two modes with $k=\pm k_{min}$ survive. Then, from
Eqs. (\ref{2.22}) and (\ref{2.23}),
\begin{equation}
\label{2.24} \langle \epsilon \rangle_{st}=
A_{\epsilon}(\widetilde{\delta L})^{-1}, \quad A_{\epsilon}=
\frac{2}{N_{x,h}L_{c}},
\end{equation}
where it has been used that near the instability it is $N \simeq
N_{x,h} L_{c}$.  This equation shows that a renormalization by
fluctuations of the total energy of the HCS takes place as the
clustering instability is approached. In fact, it is seen that the
energy of the system becomes larger than the prediction based on
Haff's law. When interpreting this theoretical prediction, it must
be kept in mind the the approximation carried out here only holds
while the fluctuations are small as compared with their average
values far from the instability. This requires, in particular, that
the right hand side of Eq.\ (\ref{2.24}) be small.

\section{DSMC method results}
\label{s3} The simulation results to be reported in the following
have been obtained by using the DSMC method, partitioning the system
in parallel layers as described above. The simulations started with
a configuration in which the particles were homogeneously
distributed along the length $L_{x}$ of the system and their
velocities obeyed a Maxwellian distribution. In order to increase
the number of statistical averages and avoid the technical problems
inherent to the continuous cooling of the HCS, the steady
representation of the latter was used. This consists in a change in
the time scale that implies a modification of the dynamics, but does
not involve any internal property of the system. This dynamics leads
to a steady state whose properties are exactly mapped into those of
the HCS. The details of the method have been extensively analyzed
elsewhere \cite{BRyM04} and will be not repeated here. Then, the
system is allowed to evolve until the steady state is reached. It is
important to verify that the system has actually reached
stationarity, since the time required to reach it increases very
fast as the instability is approached. For instance, for
$\alpha=0.9$, the system becomes stationary after about $100$
collisions per particle for $L_{x}/L_{c}=0.88$, while it needs of
the order of $2000$ collisions per particle for $L_{x}/L_{c}=0.997$.
Once the system is in the steady state, the different macroscopic
properties of interest have been computed by taking averages of the
microscopic values. In addition, in all the results reported here,
the data have also been averaged over $1000$ independent
trajectories.

The height of the cells used in the simulations has been $\Delta x =
0.04 \ell_{0}$, significantly smaller than the mean free path
$\lambda=0.225 \ell_{0}$, as required by the DSMC method. The number
of particles per cell was $100$, so $N_{x,h} \equiv N/L_{x}=2500
\ell_{0}^{-1}$. The number of cells to be considered in each case
was determined from the the above figures and the value of $L_{x}$
being simulated. It is worth to stress that the number of particles
does not affect the effective dynamics used in the DSMC that
remains, by definition, that of an $N$ particle system in the low
density (Boltzmann) limit \cite{Bi94}.

For a given value of $\alpha$, the series of simulations were
started with a system size $L_{x}$ significantly smaller than the
critical values predicted by Eq.\ (\ref{1.2}), namely around $0.8$
times that value. From there, the size of the system was increased
systematically and its properties studied in each case. Special
attention was given to verify the homogeneity of the hydrodynamic
properties of the system. Once the formation of velocity vortices
was observed, the series was stopped since this is the indication
that the size is larger than the critical length $L_{c}$.

\begin{figure}
\centering
\includegraphics[angle=-90,scale=0.45]{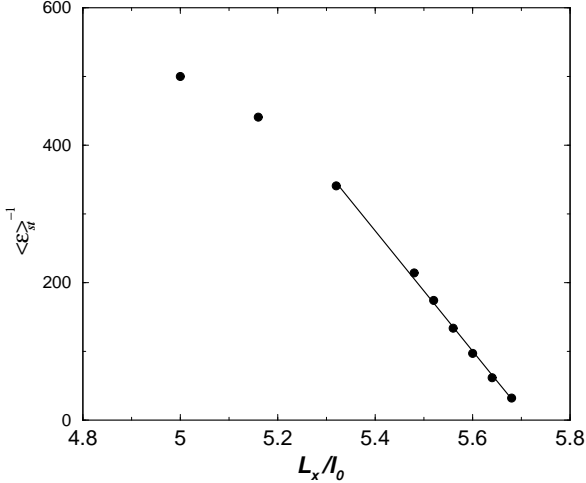}
\caption{Dimensionless quantity $\langle \epsilon
\rangle_{st}^{-1}$, defined in the main text, as a function of the
size of the system $L_{x}$, measured in units of $\ell_{0} \equiv
(n_{h} \sigma^{2})^{-1}$, for $\alpha=0.9$. The circles are the DSMC
results and the solid line a linear fit near the clustering
instability.} \label{fig1}
\end{figure}

\begin{figure}
\centering
\includegraphics[angle=-90,scale=0.45]{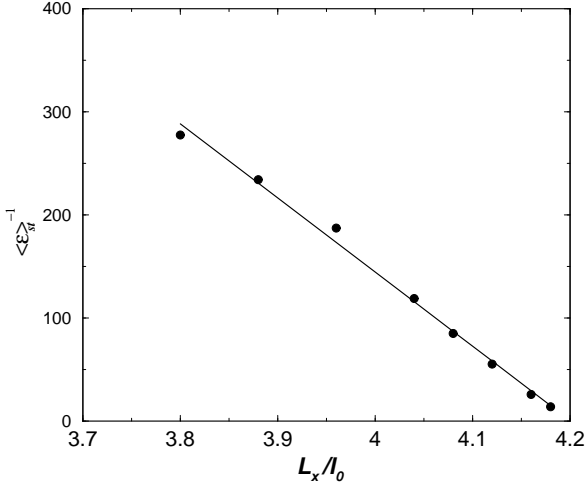}
\caption{The same as in Fig. \ref{fig1}, but for a system of
particles with $\alpha = 0.8$.} \label{fig2}
\end{figure}

In Figs. \ref{fig1} and \ref{fig2} the quantity $\langle \epsilon
\rangle_{st}^{-1}$ is plotted as a function of $\Lambda_{x}=
L_{x}/\ell_{0}$ for $\alpha=0.9$ and $\alpha=0.8$, respectively. In
both cases, an increase of the average energy relative to the Haff's
law prediction is observed as the size of the system is decreased
approaching the critical value. It must be noted, however, that this
increase remains quantitatively small, namely below $10 \%$ in all
the parameter region studied.

If the theoretical prediction in Eq.\ (\ref{2.24}) is verified, the
above plot must lead to a straight line, and that is in fact was is
obtained when $L_{x}$ increases approaching the critical length.
This confirms the exponent $-1$ in Eq.\ (\ref{2.24}). From the slope
and the ordinate at the origin, the simulation values of the
critical size $L_{c}$ and the critical amplitude $A_{E}$ are
obtained. These fitting parameters as well as the theoretical
predictions for them are given in Table \ref{table1}. The
theoretical prediction for the critical length has been obtained by
means of Eq.\ (\ref{1.2}) and the expressions for the shear
viscosity and the cooling rate in the first Sonine approximation
given in \cite{BDKyS98}. It is observed that the agreement between
the theory and the simulation results for the critical size $L_{c}$
is really good. This is consistent with the well established result
that the first Sonine approximation provides good values for the
shear viscosity and the cooling rate of a dilute granular gas in the
range of values of $\alpha$ considered here.

On the other hand, the comparison between the theoretical prediction
and the DSMC result for the critical amplitude of the energy
$A_{\epsilon}$ is not so good. In Table \ref{table1} the measured
values for $N_{x}L_{c}A_{\epsilon}$ is given. According to Eq.\
(\ref{2.24}) it should be equal to a constant, independent of the
inelasticity, namely $2$. Instead, although the order of magnitud is
accurately  predicted, definitely larger values are obtained
showing, in addition, a relevant dependence on $\alpha$. Something
about the possible origins of this discrepancy and the improvement
of the theory will be said at the end of the paper.

\begin{table}[t]
\caption{Theoretical prediction, $L_{c}^{(t)}$, and values obtained
from the simulation data for the average energy,
$L_{c}^{(\epsilon)}$, and from the second moment of the
fluctuations, $L_{c}^{(\sigma)}$ for the critical length
characterizing the clustering instability of a dilute
three-dimensional granular gas, as a function of the coefficient of
restitution $\alpha$. In all cases, $L_{c}$ is measured in units of
$\ell_{0} = (n_{h} \sigma^{2})^{-1}$. The parameters reported in the
last two columns involve the critical amplitudes $A_{\epsilon}$ and
$A_{\sigma}$ and should be both equal to $2$, according to the
theoretical predictions.} \centering \label{table1}
\begin{tabular}{llllll}
\hline\noalign{\smallskip} $ \alpha $ & $ L_{c}^{(t)}$ & $
L_{c}^{(s, \epsilon)} $& $L_{c}^{s, \sigma}$ & $ L_{c}^{(\epsilon
)}N_{x}A_{\epsilon}^{(s)}$
& $(A_{\sigma}^{(s)} N_{x} L_{c}^{(\sigma)})^{2}$ \\
[3pt] \tableheadseprule\noalign{\smallskip}
0.95 & 7.86 & 7.93  & 7.91 & 2.66 & 2.08 \\
0.90 & 5.70 & 5.71  & 5.71 & 2.72 & 2.78 \\
0.85 & 4.77 & 4.75  & 4.75 & 2.82 & 2.43 \\
0.80 & 4.23 & 4.19  & 4.19 & 2.85 & 2.68 \\
0.75 & 3.88 & 3.82  & 3.82 & 3.35 & 2.88 \\
\noalign{\smallskip}\hline
\end{tabular}
\end{table}

\section{Critical energy fluctuations}
\label{s4} 
\begin{figure}
\centering
\includegraphics[angle=-90,scale=0.45]{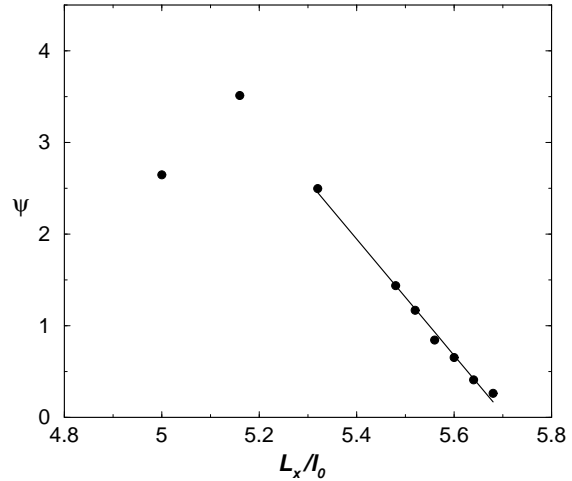}
\caption{Plot of the DSMC results (circles) for the dimensionless
quantity defined in Eq.\ (\protect{\ref{4.10}}) as a function of the
size of the system $L_{x}$, for $\alpha=0.9$. The solid line is a
linear fit near the clustering instability.} \label{fig3}
\end{figure}

\begin{figure}
\centering
\includegraphics[angle=-90,scale=0.45]{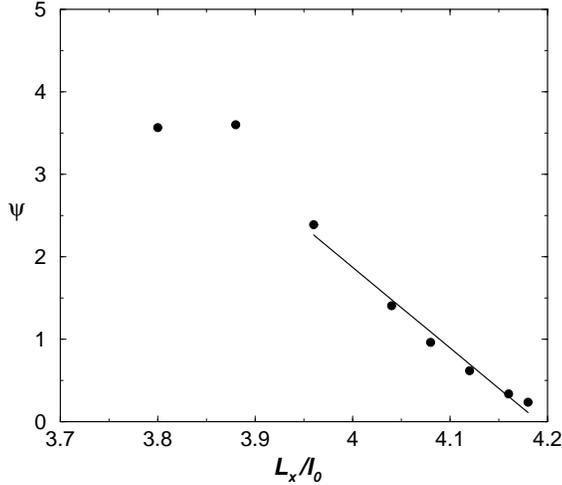}
\caption{The same as in Fig. 3, but for $\alpha=0.8$.} \label{fig4}
\end{figure}

Consider next the fluctuations around its average value
of the total energy of the system, again in the threshold of the
instability. Since it has been shown above that the average value of
the energy differs from the Haff law prediction, let us define
\begin{equation}
\label{4.1} \delta^{\prime} \epsilon (s) \equiv \epsilon (s) - <
\epsilon>_{st} = \frac{\widetilde{E}(s) - \langle \widetilde{E}(s)
\rangle }{E_{H}(s)}\, ,
\end{equation}
\begin{equation}\label{4.2}
\delta^{\prime} \overline{\omega}(s) \equiv \overline{\omega}(s) -
\langle \overline{\omega}(s) \rangle_{st}\, ,
\end{equation}
with
\begin{equation}
\label{4.3} \overline{\omega}(s) \equiv \frac{2}{\Lambda_{x}^{2} }
\sum_{k} |\omega_{k, \perp}(s)|^{2}.
\end{equation}
Equation (\ref{2.22}) shows that $<\overline{\omega}>_{st}
=<\epsilon>_{st}$ and, therefore, Eq.\ (\ref{2.21}) is equivalent to
\begin{equation}
\label{4.4} \frac{ d}{d s} \delta^{\prime} \epsilon (s) = -
\frac{\zeta^{\ast}}{2} \left[  \delta^{\prime} \epsilon(s) -
\delta^{\prime} \overline{\omega}(s) \right].
\end{equation}
Taking into account that cumulants of ${\bm \omega}_{k,\perp}$ of
order higher than two vanish since ${\bm \xi}_{\perp}^{(\omega}(s)$
is assumed to be Gaussian, it follows from Eqs.\ (\ref{2.14}) and
(\ref{2.15}) that
\begin{equation}
\label{4.5} < \delta^{\prime} \overline{\omega}(s) \delta^{\prime}
\overline{\omega} (s^{\prime}) > =\frac{2}{N_{x,h}^{2}L_{c}^{2}}
e^{-2(s-s^{\prime}) \zeta^{*} \widetilde{\delta L}}
(\widetilde{\delta L})^{-2}.
\end{equation}
Upon deriving this equation, it has been used that only modes with
${\bm k}$ in the direction of the $x$ axis can be excited, due to
the way in which the simulations are carried out. The solution of
Eq.\ (\ref{4.4}), once the initial value has been forgotten, is
\begin{equation}
\label{4.6} \delta^{\prime}\epsilon (s) = \frac{\zeta^{*}}{2}
\int_{-\infty}^{s} d s_{1} e^{ -\zeta^{*}(s-s_{1})/2} \delta
^{\prime} \overline{\omega} (s_{1}).
\end{equation}
Then, by means of Eq.\ (\ref{4.5}) it follows that
\begin{eqnarray}
\label{4.7} <\delta^{\prime} \epsilon (s) \delta^{\prime} \epsilon
(s^{\prime})> & = &  \frac{2}{N_{x,h}^{2}L_{c}^{2}}
e^{-2(s-s^{\prime}) \zeta^{*} \widetilde{\delta L}}
(\widetilde{\delta L})^{-2} \nonumber \\
& = & < \delta^{\prime} \overline{\omega}(s) \delta^{\prime}
\overline{\omega} (s^{\prime})
> ,
\end{eqnarray}
$s \geq s^{\prime} \gg 1$. Terms involving powers of
$\widetilde{\delta L}$ larger than $-2$ have been consistently
neglected in this expression. Thus both the total energy
fluctuations and the fluctuations of the energy associated with the
transversal modes decay with a characteristic time $( 2 \zeta^{*}
\widetilde{\delta L})^{-1}$, indicating a divergent behavior of the
relaxation time as $\widetilde{\delta L}^{-1}$. Also, the second
moment of the energy fluctuations is predicted to diverge. For
$s=s^{\prime}$, Eq. (\ref{4.7}) leads to the stationary value
\begin{equation}
\label{4.8} < (\delta^{\prime} \epsilon )^{2}
>_{st}= A_{\sigma}^{2} \widetilde{\delta L}^{-2}, \quad
A_{\sigma}= \frac{\sqrt{2}}{N_{x,h} L_{c}}\, .
\end{equation}
It is
\begin{equation}
\label{4.9} \sigma_{E}^{2} \equiv \frac{<( \widetilde{E}(s)-<
\widetilde{E}(s)>)^{2}>}{< \widetilde{E}(s)>}  \simeq <
(\delta^{\prime} \epsilon )^{2}
>.
\end{equation}
To compare with the DSMC method results, it must be realized that
Eq.\ (\ref{4.8}) gives the moment of the energy fluctuations that
dominate near the instability, but in a region where the
fluctuations are still small. Then, what has been done is to measure
$\sigma_{E}^{2}$ as a function of $L_{x}$, starting from values
quite smaller than $L_{c}$. There, the value of $\sigma_{E}$   is
not affected by the presence of the instability, and $N
\sigma_{E}^{2}$ is independent of $L_{x}$ \cite{BGMyR04}. It will be
denoted by $(N \sigma_{E}^{2})_{h}$. Then, the quantity considered
has been
\begin{equation}
\label{4.10} \psi \equiv \left[ \frac{N_{x,h} L_{c}
\sigma_{E}^{2}}{(N \sigma_{E}^{2})_{h}} -1 \right]^{-1/2}.
\end{equation}
The measured values for this quantity are plotted as a function of
$L_{x}$ in Figs.\ \ref{fig3} and \ref{fig4}, for $\alpha=0.9$ and
$\alpha =0.8$, respectively. As predicted by Eq.\ (\ref{4.8}), a
linear behavior is observed as the critical size is approached.
Although the fit is reasonably good for all the values of $\alpha$
reported here, it becomes clearly worse as the inelasticity
increases. Moreover,  the `critical' region, identified by the
values of $L_{x}$ for which the predicted qualitative critical
behavior is actually observed, is smaller for the second moment of
the energy than for its average. This can be verified by comparing,
for instance, Figs.\ \ref{fig2} and \ref{fig4}. In contrast to what
happened with the average energy, the growth exhibited by the
fluctuations is quite fast. In the region plotted in Figs.
\ref{fig3} and \ref{fig4}, the second moment $\sigma_{E}$ increases
more than one order of magnitude.

>From the slope and ordinate in the origin of the fitting straight
line, the simulation values of the critical length $L_{c}$ and
amplitude $A_{\sigma}$ are directly derived. The values obtained in
this way are also included in Table \ref{table1}. Again, the results
for the critical length $L_{c}$ are in excellent agreement with the
theoretical predictions. Moreover, the fact that the same values
follow from the measurements of both the average energy and the
second moments provides a test of the internal consistency of the
theory. With regard to the critical amplitude $A_{\sigma}$,
significant deviations from the the theoretical predictions are also
found in this case, although they are smaller than for
$A_{\epsilon}$. In fact, for the most elastic system considered,
$\alpha=0.95$, there is a good agreement. The most relevant
disagreement  between theory and simulations is the definite
dependence on inelasticity of the parameter combination given in
Table \ref{table1} shown by the latter, while the former predicts a
constant value (namely, $2$).

It can be wondered which is the aspect of the theoretical approach
developed here that should be modified in order to improve the
accuracy of the predicted  expressions for the critical amplitudes.
Of course, a first important limitation of the theory is its
restriction  to the almost elastic limit, implied by the use of the
Landau fluctuation-dissipation relations. But it must be realized
that up to now those expressions have not been generalized for
non-conservative interactions. A more modest step in this direction
should be to include in Eq.\ (\ref{2.18}) the intrinsic noise term
associated with the cooling rate, as already indicated.

In refs. \cite{BGMyR05} and \cite{BDGyM06}, a scaling property of
the probability distribution of the energy fluctuations in the
threshold of the clustering instability of a two-dimensional
granular fluid was identified. Moreover, the scaling function was
very well fitted by the same expression as found in several
equilibrium and non-equilibrium molecular systems
\cite{BHyP98,BCFetal00}. The possible existence of a similar scaling
for the three-dimensional granular gas considered here has also been
investigated, finding similar results.

\begin{acknowledgements}
This research was supported by the Ministerio de Educaci\'{o}n y
Cienc\'{\i}a (Spain) through Grant No. FIS2005-01398 (partially
financed by FEDER funds).
\end{acknowledgements}

\end{document}